\documentclass[aps,twocolumn,superscriptaddress,showpacs,showkeys,longbibliography]{revtex4-1}
\usepackage{graphics,graphicx,dcolumn,bm,fleqn,epic,eepic,float,epsfig}
\usepackage{amssymb,amsmath,multirow,rotate,color,float}
\usepackage{epstopdf}
\usepackage{times,natbib}
\usepackage{color}
\usepackage{soul}                           
\definecolor{red}{rgb}{1,0,0}
\definecolor{green}{rgb}{0,1,0}
\definecolor{blue}{rgb}{0,0,0} 
\definecolor{gray}{rgb}{0.9,0.9,0.9}

\newcommand{\xxb}[1]{\textcolor{blue}{#1}} 

\begin{document}
\title{Directed percolation in aerodynamics: 
       resolving laminar separation bubble on airfoils}

\author{Dominik Traphan}
\affiliation{ForWind, Institute of Physics, University of Oldenburg,
	K\"upkersweg 70, 26129 Oldenburg, Germany}
\author{Tom T. B. Wester}
\affiliation{ForWind, Institute of Physics, University of Oldenburg,
	K\"upkersweg 70, 26129 Oldenburg, Germany}
\author{Gerd G\"ulker}
\affiliation{ForWind, Institute of Physics, University of Oldenburg,
	K\"upkersweg 70, 26129 Oldenburg, Germany}
\author{Joachim Peinke}
\affiliation{ForWind, Institute of Physics, University of Oldenburg,
	K\"upkersweg 70, 26129 Oldenburg, Germany}
\author{Pedro G.~Lind}
\affiliation{Institute of Physics, University of Osnabr\"uck,
	Barbarastrasse 7, 49076 Osnabr\"uck, Germany}

\date{\today}

\begin{abstract}
	In nature, phase transitions prevail amongst inherently different systems, while frequently showing a universal behavior at their critical point. As a fundamental phenomenon of fluid mechanics, recent studies suggested laminar-turbulent transition belonging to the universality class of directed percolation. Beyond, no indication was yet found that directed percolation is encountered in technical relevant fluid mechanics. Here, we present first evidence that the onset of a laminar separation bubble on an airfoil can be well characterized employing the directed percolation model on high fidelity particle image velocimetry data. In an extensive analysis, we show that the obtained critical exponents are robust against parameter fluctuations, namely threshold of turbulence intensity that distinguishes between ambient flow and laminar separation bubble. Our findings indicate a comprehensive significance of percolation models in fluid mechanics beyond fundamental flow phenomena, in particular, it enables the precise determination of the transition point of the laminar separation bubble. This opens a broad variety of new fields of application, ranging from experimental airfoil aerodynamics to computational fluid dynamics.
\end{abstract}



\maketitle

\section{Introduction}

Three decades ago, Pomeau described the flow of a fluid as a 
collection of oscillators that interact with each other. 
When observing that \cite{pomeau1986} ``each oscillator if in a turbulent 
state may either relax spontaneously toward its quiescent state or 
contaminate its neighbors'' he concluded that ``this is precisely the 
definition of the process called “directed percolation” in statistical 
physics'' and therefore raised the possibility of laminar-turbulent
transition belonging to the same universality class as directed
percolation (DP).

The idea is quite reasonable in the sense that transition from laminar 
to turbulent flow \xxb{(hereafter referred to as transition) can be described via} the so-called spatio-temporal 
intermittency \cite{chate1987,rupp2003}. While since then several simulations 
have supported Pomeau's conjecture, only in the last \xxb{few years} it was 
possible to provide experimental evidence \cite{pomeau2016}, due to 
novel possibilities of extracting accurate measurements from a turbulent 
flow with sufficient spatial and temporal resolution.
\xxb{Recent studies approached transition from different angles by means of low-order models \cite{eckhardtDE2,sipos2011,barkley2011}, sophisticated simulations \cite{shih2016,kreilos2014,avila2013}, as well as experiments \cite{lemoult2015,sano2016,barkley2015}. They concordantly indicate non-equilibrium phase transition occurring in basic shear flows, i.e. pipe, channel and Couette flows.}
While investigating the hypothesis of transition belonging to the universality 
class of directed percolation, they treat transition into turbulence as a fundamental phenomenon. 
Up to now, directed percolation has not been used in flows with a direct aim of
helping to solve specific needs in applied sciences that deal with turbulence by extension.
\begin{figure}[t]
\begin{center}
\includegraphics[width=0.88\columnwidth]{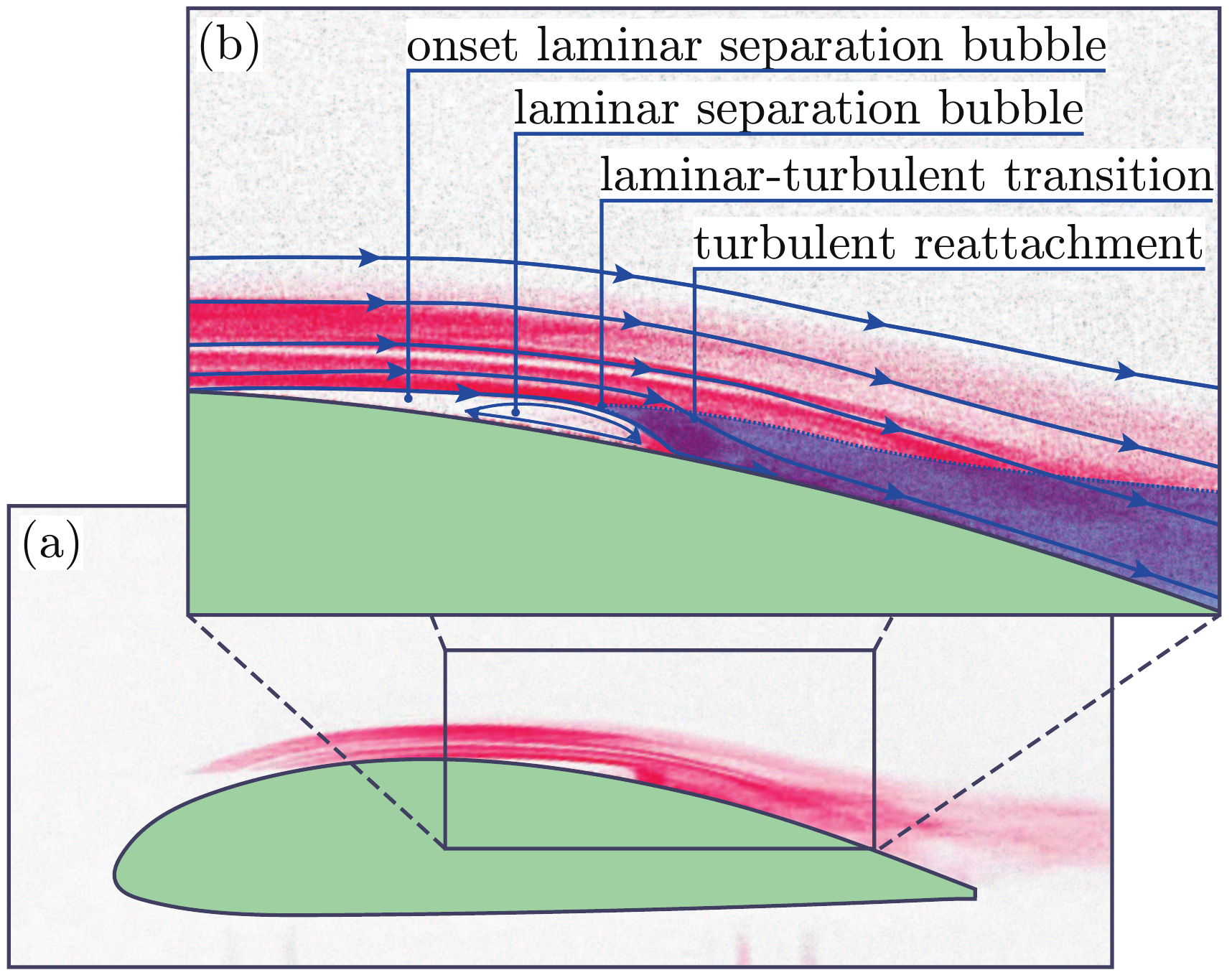}
\caption{\protect
(a)
Photograph of an LSB on airfoil CK220. Laser light sheet illuminates the particle seeded inflow from right-hand side and perpendicular to the airfoil's surface resulting in a shadow at the leading edge. Inflow is coming from left-hand side. LSB can be identified as a region without smoke between the airfoil's surface and the ambient flow. False colors are used for better visualization.
(b)		
Labeled zoom, emphasizing the flow topology at the LSB. LSB occurs just behind the airfoil's thickest cross section, where streamlines (in blue) separate from the surface. Further downstream, when LSB is at its maximum expansion, the laminar shear layer destabilizes resulting in transition to turbulence. High momentum perpendicular to the airfoil's surface contained in the shear layer enables reattachment and the formation of a turbulent boundary layer.}
\label{fig01}
\end{center}
\end{figure}

Here, we argue that the applicability of directed percolation model can be extended from fundamental fluid dynamics to practical aerodynamics relevant for engineering problems and, thus, to a more generally valid concept. More specifically, we provide evidence that directed percolation is capable of characterizing the onset of a laminar separation bubble (LSB) on the suction side of an airfoil (see Fig.~\ref{fig01}) \cite{Horton1968,Omeara1987}.
\xxb{This is of particular importance since the aerodynamic functionality of an airfoil depends sensitively on such transitions like the LSB. Furthermore, the precise determination of the onset of transition into LSB as well as transition in general are still an open problem in aerodynamics. So far, there are no methods available that reliably capture the nonlinear nature of an LSB. In this way, DP is potentially valid for nonlinear flow phenomena encountered in the class of external flows over curved surfaces, where the flow becomes more and more unstable as it is advected downstream. Based on our findings, we discuss how DP properties can also be of use in applications, ranging from computational fluid dynamics (CFD), through the design process of airfoils, to maintenance of rotor blades in wind turbines.} 

\xxb{The paper is structured as follows. At first, the experimental setup is described in Sec.~\ref{sec:experiment}. This includes an airfoil subjected to laminar inflow and the optical flow visualization method. In Section~\ref{sec:methods}, a procedure is explained how to map the measured flow field into binary cells, either laminar or turbulent, which enable for estimating a critical Reynolds number identifying the onset of LSB. Characteristic exponents at this critical value are determined in Sec.~\ref{sec:exponents} and discussed subsequently in Sec.~\ref{sec:discussion}. Finally, Section~\ref{sec:conclusions} concludes the paper.}

\section{The Experiment}
\label{sec:experiment}

The phenomenon of LSB can be qualitatively visualized by means of a laser 
illuminated smoke photography (see Fig.~\ref{fig01}). 
The seeded flow is coming from left-hand side, separates just after 
the airfoil's thickest cross section, destabilizes and undergoes transition 
resulting in reattachment of the turbulent boundary layer. 
The recirculation zone between the airfoil's surface and the shear layer 
is the one usually addressed as LSB. 
In Figures \ref{fig01}a and \ref{fig01}b, the LSB appears as the smoke 
free region above the airfoil. 
While transition, taking place within the shear layer between LSB and 
free flow, is three-dimensional, the LSB's onset happens in a
linearly stable laminar flow region \xxb{whose boundary layer has a thickness small compared to the dimensions of the LSB}.
Therefore, transition into LSB can be approximated locally as a quasi
two-dimensional process.

\begin{figure}[t]
\begin{center}
\includegraphics[width=0.9\columnwidth]{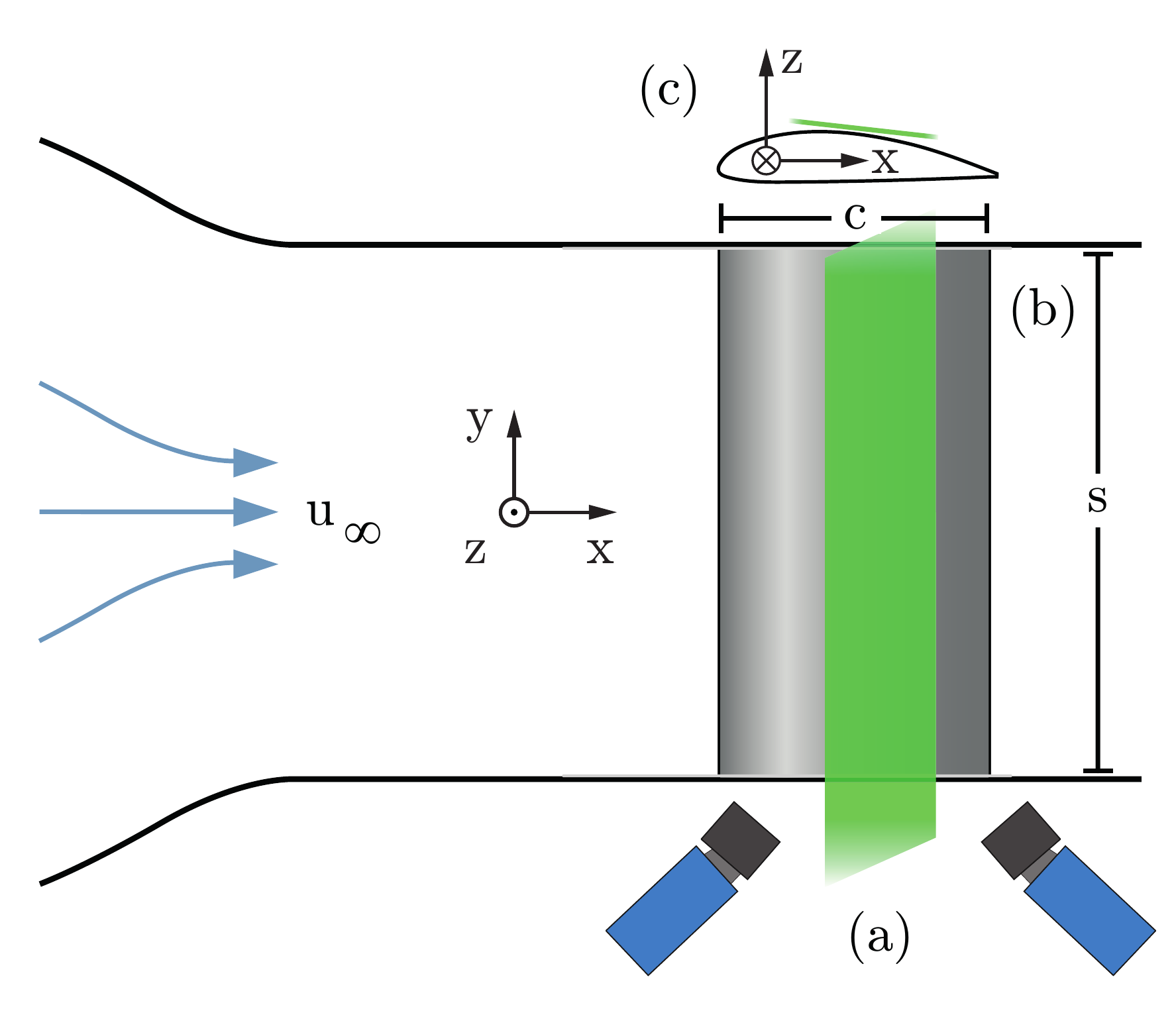}
\caption{\protect
  \xxb{Experimental setup consisting of (a) a PIV system and (b) a CK220 airfoil. Laminar inflow of $u_{\infty} = 11~{\rm  m/s}$ approaches airfoil from left-hand side forming a laminar separation bubble (LSB) on the suction side. The light-sheet (in green) is adjusted (c) tangentially to the surface approximately at the location where LSB is observed. The coordinate system has its origin at the mid-span leading edge and defines the main flow direction as x-direction, spanwise as y-direction and normal to wind tunnel wall as z-direction. Chord length and span of the airfoil are denoted $c$ and $s$, respectively.}
}
\label{fig02}
\end{center}
\end{figure}

The formation of an LSB is a highly unsteady process which emerges 
stochastically over a certain region on airfoils. In order 
to properly investigate this complex and delicate flow topology, 
it is crucial to use non intrusive methods featuring high spatio-temporal resolution. Therefore, stereoscopic high-speed particle image velocimetry (HSPIV) is used to visualize an LSB on the suction side 
of a CK220 airfoil in the wind tunnel. 
A schematic representation of the experiment is given in 
Fig.~\ref{fig02}. 
\xxb{%
The experimental setup consists of two Phantom Miro M320S 
high-speed cameras and a Litron LDY303 laser. This enables HSPIV 
measurements with a recording frequency of 2,000 velocity fields per 
second 
\xxb{and a recording length of $T \approx 3~{\rm s}$} 
at reduced resolution of $896\times792~{\rm px}^2$.}
For a measurement region of 
$\Delta x\times\Delta y=40\times40~{\rm mm}^{2}$, 
where $x$ and $y$ are in chordwise, respectively spanwise, direction, 
a spatial resolution of \xxb{$dx=dy<0.4~{\rm mm}$} is obtained with 
a sufficient accuracy (stereo residue below $0.5~{\rm px}$).
This resolution corresponds to 
\xxb{$dx/c < 2\times 10^{-3}$} at given airfoil dimension of 
$c\times s = 220\times 250~{\rm mm}^{2}$, where $c$ denotes the 
airfoil's chord length and $s$ is its span.

As illustrated in Fig. \ref{fig02}, the light sheet is adjusted 
tangentially to the airfoil's surface approximately at
the onset of the LSB at a distance \xxb{$dz/c < 3\times 10^{-3}$ or $dz = 0.5~{\rm mm}$, respectively}.
Due to the airfoil's curvature, this distance varies slightly, by less than $5\%$ of the LSB's thickness. The onset of LSB is thus captured properly. 
At the same time, the measurement region covers the LSB completely.

\xxb{In the present experiment, the LSB appears at a global Reynolds number of ${\rm Re}_{{\rm chord}} = 160,000$ and an angle of attack of $\alpha = 5^{\circ}$. ${\rm Re}_{{\rm chord}}$ is obtained with respect to the chord length and a free stream velocity of $u_{\infty} = 11~{\rm m/s}$. In this way, the ambient conditions of the experiment are described in general.} Moreover, any transition process is strongly dependent on detailed flow conditions, such as free stream properties. To minimize this impact, the experiments are performed in a wind tunnel with inlet $250\times 250~{\rm mm}^{2}$, length $2000~{\rm mm}$ and a closed test section with a low turbulence intensity of \xxb{${\rm TI} < 0.004$.}

\section{From particle image velocimetry data to laminar and turbulent states}\label{sec:methods}

\begin{figure*}[t]
	\begin{center}
		\includegraphics[width=\columnwidth]{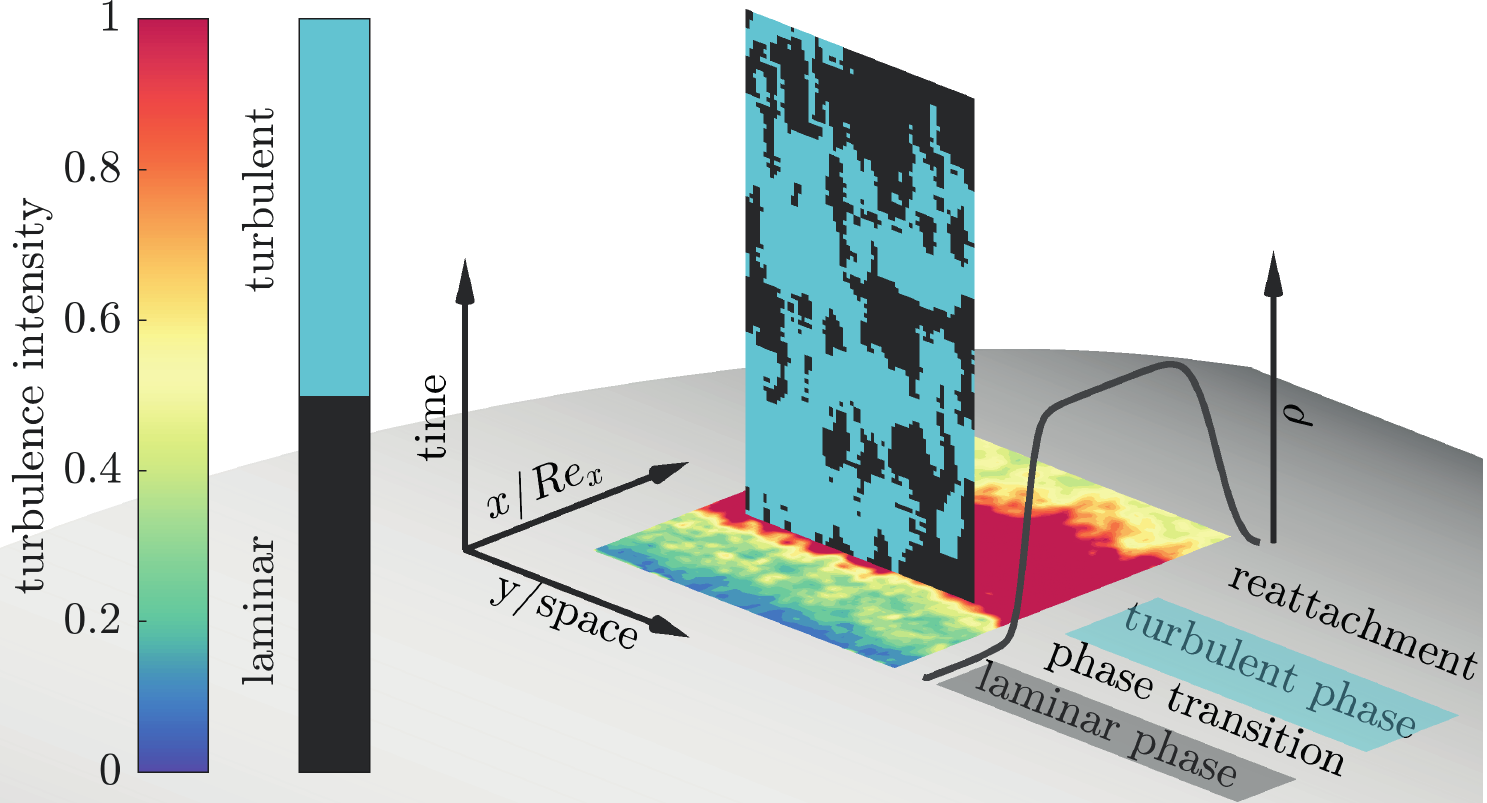}
		\caption{\protect
			Illustration of measured and computed quantities: The plane of space and control parameter (${\rm Re}_x$) is formed by the airfoil's spanwise and chordwise direction, where ${\rm Re}_{x} = {\rm Re}(x, u_{\infty})$ is based on the distance $x$ from the leading edge and the inflow velocity $u_{\infty}$. ${\rm Re}_{x}$ provides a measure for the flow's likelihood to undergo transition from laminar flow into an LSB and, therefore, serves as a control parameter. The presence of LSB is indicated by means of turbulence intensity (${\rm TI}$) exceeding a certain threshold denoted as turbulent phase. The instantaneous turbulence intensity is derived from 10 temporally consecutive velocity fields at one respective position ${\rm Re}_{x}$ in a plane crossing the laminar separation bubble. Phase transition from laminar flow into LSB happens once turbulent clusters merge significantly. At this critical point, ${\rm Re}_{\rm c}$, characteristic turbulent clusters occur, depicted as an evolution in time (vertical plane). Averaging over time at each position ${\rm Re}_{x}$, the fraction of turbulent cells $\rho$ is shown next to the plane of TI. \xxb{Following $\rho$ in streamwise direction, phase transition from laminar boundary layer flow into the LSB and subsequent reattachment are obvious. The first phase transition represents the LSB's onset which is important for our work.}}
		\label{fig03}
	\end{center}
\end{figure*}

Following the objective of this study, the LSB's onset needs 
to be derived from experimental data. HSPIV results provide instantaneous 
velocity fields of the whole LSB \xxb{and the flow upstream as well as} downstream of 
the LSB. Characteristic differences between those three regions are reflected by 
velocity magnitude $u$ and the fluctuations $u^{\prime}$. Both quantities define the turbulence intensity ${\rm TI}$.
The value of ${\rm TI}$ is usually derived locally from the whole data set losing its temporal information, whereas directed percolation analysis is based on the time-resolved evolution of laminar and turbulent regions. \xxb{Therefore, we define the ${\rm TI}$ at the time $t$ as the ${\rm TI}$ computed locally at the position $(x,y)$ from 10 consecutive velocity fields in the following manner:
	\begin{subequations}
		\begin{eqnarray}
		\langle u(x,y,t) \rangle  \equiv \langle u_{t} \rangle =  \frac{1}{10} \sum_{t'=t}^{t+9} u(x,y,t'),~~~~~~~~~~~~~~~~~~~\label{sub1}\\
		{\rm TI}(x,y,t) =  \sqrt{ \frac{1}{10 {\langle u_{t} \rangle}^2} \sum_{t'=t}^{t+9} (u(x,y,t') -\langle u_{t} \rangle)^{2} } .\ \ \ \label{sub2}
		\end{eqnarray}
	\end{subequations}
}

\xxb{The general mapping procedure into laminar and turbulent states is visualized in Fig.~\ref{fig03}. The flow is moving along the airfoil's surface (gray) in the x-direction which is chordwise. In the horizontal color-coded plane, the LSB can be clearly identified as a high level of turbulence 
intensity. Mainly a velocity magnitude close to zero within the LSB contributes to the sharp rise of $\rm TI$, while velocity fluctuations increase only slightly. 
It is important to notice that the term ``laminar separation bubble'' is misleading at this point. In the context of the present work, the LSB is revealed by a turbulent flow region, whereas it is referred to as laminar in airfoil aerodynamics. This is due to the fact that velocity fluctuations within the LSB are small compared to fluctuations present in a fully turbulent boundary layer. With that in mind, we use a threshold value of the turbulence intensity ${\rm TI}_{{\rm th}}$ to distinguish between laminar and turbulent  regions, called clusters. Each measurement point of the complete dataset is set to 0 if ${\rm TI}(x,y,t) < {\rm TI}_{\rm th}$ (laminar phase) or 1 if ${\rm TI}(x,y,t) \geq {\rm TI}_{\rm th}$ (turbulent phase). The turbulent phase corresponds to the LSB whereas laminar phase identifies ambient flow.}

\xxb{The onset of the LSB is accurately determined by the evolution of laminar and turbulent clusters. The distribution of clusters is evaluated at each position $x$ along the spatial and temporal dimensions $y$ and $t$, as exemplary illustrated close to the LSB's onset by the bicoloured vertical plane in Fig. \ref{fig03}.}

\section{Universal phase transition into laminar separation bubble}\label{sec:exponents}

\xxb{The flow along the airfoil is more likely to separate from the surface (onset of LSB) the further downstream from the leading edge it has been evolved due to an increasing receptivity for perturbations. In analogy to well known transition results from flat plates \cite{Schlichting2003}, this is expressed in terms of local Reynolds number ${\rm Re(x)} = {\rm Re}_{x} = x~u_{\infty}/\nu_{kin}$, where $\nu_{kin}$ denotes the kinematic viscosity. This locally changing ${\rm Re}_{x}$ represents the state of the boundary layer and  must not be confused with the fixed ${\rm Re}_{{\rm chord}}$ reflecting the experiment globally. Therefore, ${\rm Re}_{x}$ is used as the control parameter in the framework of DP for the present experiment. Notice that one HSPIV snapshot is composed of the control parameter ${\rm Re}_{x}$-axis and the spatial dimension $y$ of our percolation model. The time dimension $t$ in this percolation model corresponds to the physical time in the experiment.}

\begin{figure}[t]
	\begin{center}
		\includegraphics[width=0.95\columnwidth]{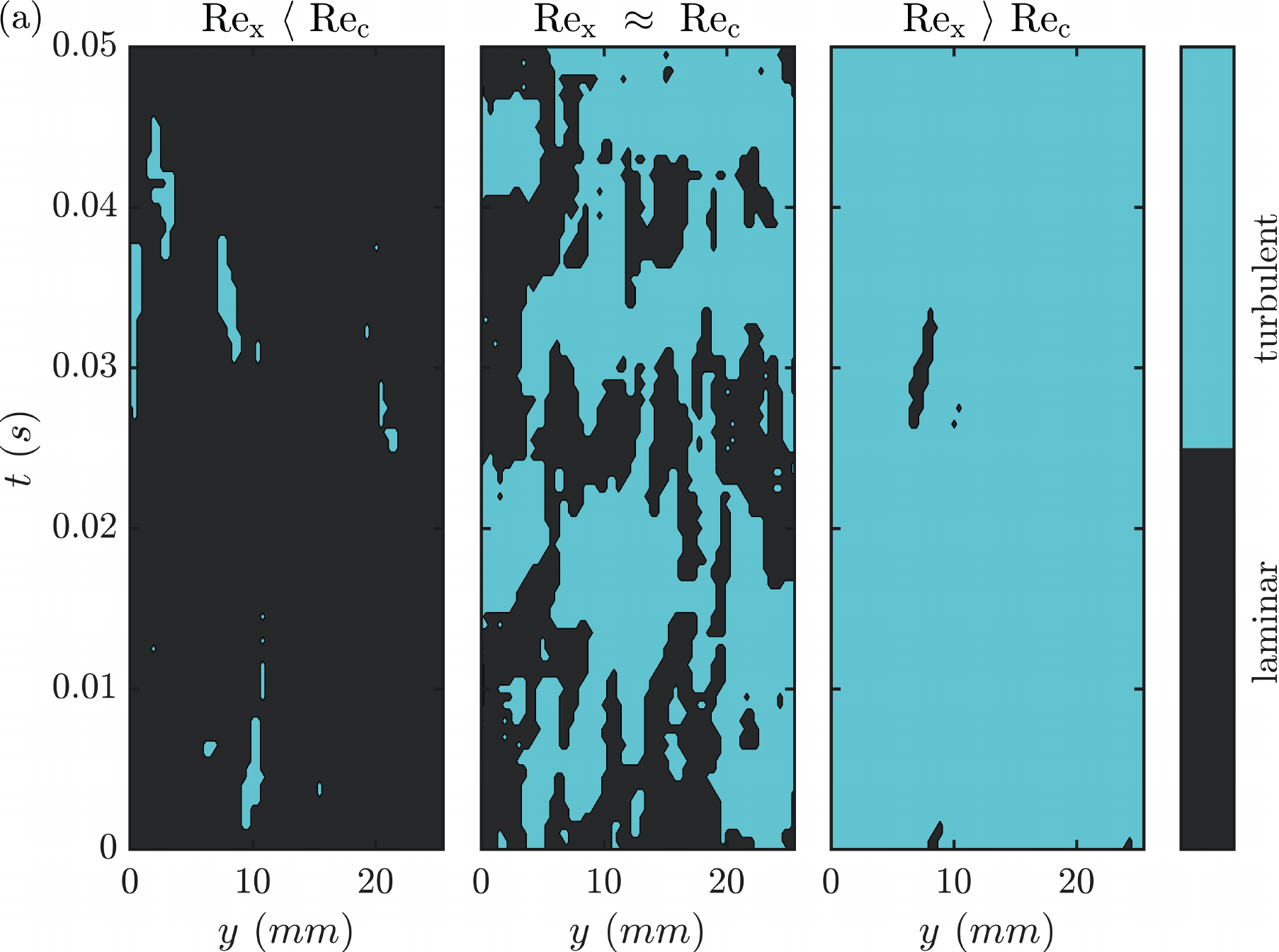}
		\includegraphics[width=0.95\columnwidth]{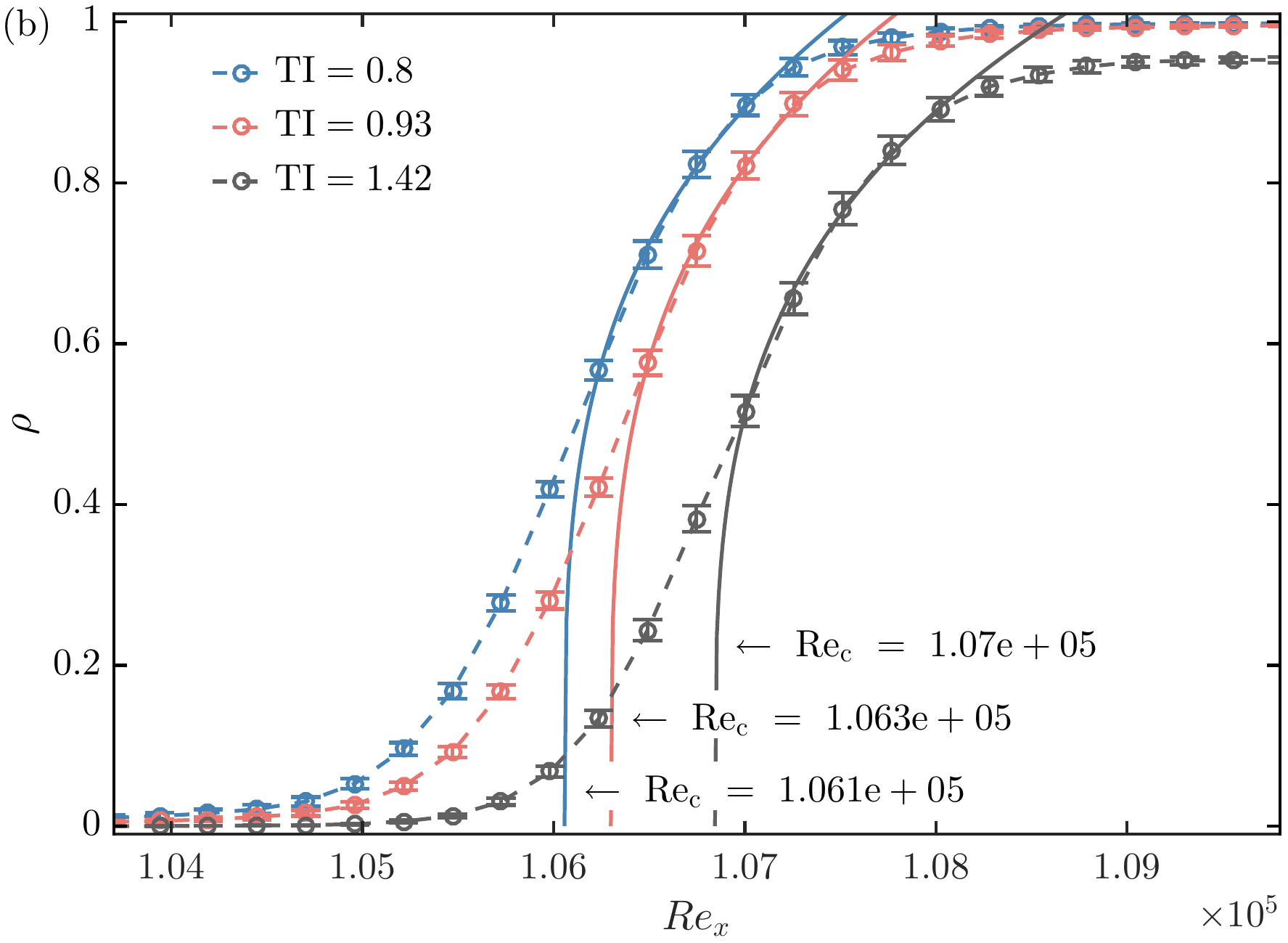}
		\caption{\protect
			\xxb{%
				(a) Three illustrations of cluster distributions in $t$ (time) and $y$ (space) derived for ${\rm TI}_{\rm th}=0.8$: 
				(left) Reynolds number below the critical value ${\rm Re}_{\rm c}$, showing a dominance of laminar clusters, 
				(right) Reynolds number above ${\rm Re}_{\rm c}$, where turbulent cells dominate, and 
				(middle) one at the computed ${\rm Re}_{\rm c}$ for which laminar and turbulent clusters span the entire system. 
			  (b) Turbulent fraction $\rho$ (circles), as a function of ${\rm Re}_{x}$, where $\rho$ is the fraction of cells with a turbulence intensity larger than a threshold ${\rm TI}_{\rm th}$. Results shown for three different threshold values. The corresponding best fits (solid lines) above the critical point cross the control parameter axis at the critical Reynolds number, $\rm Re_{\rm c}$. Experimental uncertainty of $\rho$ is estimated by the standard error of 10 subsets constituting the total dataset.}
		}
		\label{fig04}
	\end{center}
\end{figure}

\xxb{In accordance with DP, a critical value ${\rm Re}_{{\rm c}}$ of the Reynolds number ${\rm Re}_{x}$ is observed, where phase transition to LSB takes place. In general, this value is characterized by turbulent clusters merging into one infinitely connected cluster. Figure \ref{fig04}a shows three illustrative extracts containing laminar and turbulent clusters in space $y$ and time $t$. While below/above the critical Reynolds number laminar/turbulent clusters dominate, at ${\rm Re}_{{\rm c}}$ the number of laminar and turbulent clusters span the entire system in both directions.}

\xxb{The DP phase transition from the laminar boundary layer into the LSB is characterized by three critical exponents at ${\rm Re}_{{\rm c}}$. The first exponent, $\beta$, describes the critical behavior of the so-called turbulent fraction, 
\begin{equation} \label{eq:beta}
	\rho({\rm Re}_{x}) = \rho_0~\varepsilon^{\beta},
\end{equation}
as a function of the reduced Reynolds number, $\varepsilon :=({\rm Re}_{x}-{\rm Re}_{\rm c})/{\rm Re}_{\rm c}$, and a proportionality factor, $\rho_0$. The mean fraction is determined from turbulent cells over space $y$ and time $t$ for each value ${\rm Re}_{{x}}$. The other two critical exponents, $\nu_{\perp}$ and $\nu_{\parallel}$, characterize the diverging correlations of cluster sizes at the ${\rm Re}_{\rm c}$ in space and time. According to the so-called hyperscaling relation, $\mu = 2- \beta/\nu$ \cite{Takeuchi2007}, the correlation lengths are univocally expressed by the transverse and longitudinal fractal dimensions, $\mu_{\perp}$ and $\mu_{\parallel}$. These are defined as the exponents that relate the size of laminar clusters and their number, $N(L_\perp) \sim L_\perp^{-\mu_{\perp}}$ and $N(L_\parallel) \sim L_\parallel^{-\mu_{\parallel}}$ respectively, where $L_\perp$ and $L_\parallel$ represent the size of laminar clusters measured in the spatial and temporal directions, $y$ and $t$ respectively. In analogy of the  DP analysis presented by \cite{lemoult2015} and \cite{sano2016}, we estimate from our measured data first the critical Reynolds
number ${\rm Re}_{{\rm c}}$ and subsequently the set of critical exponents ($\beta, \mu_{\perp}, \mu_{\parallel} $). The results will be discussed in the following Sec. V.}

\begin{figure}[t]
	\begin{center}
		\includegraphics[clip,viewport = 0 0 535 390,width=0.95\columnwidth]{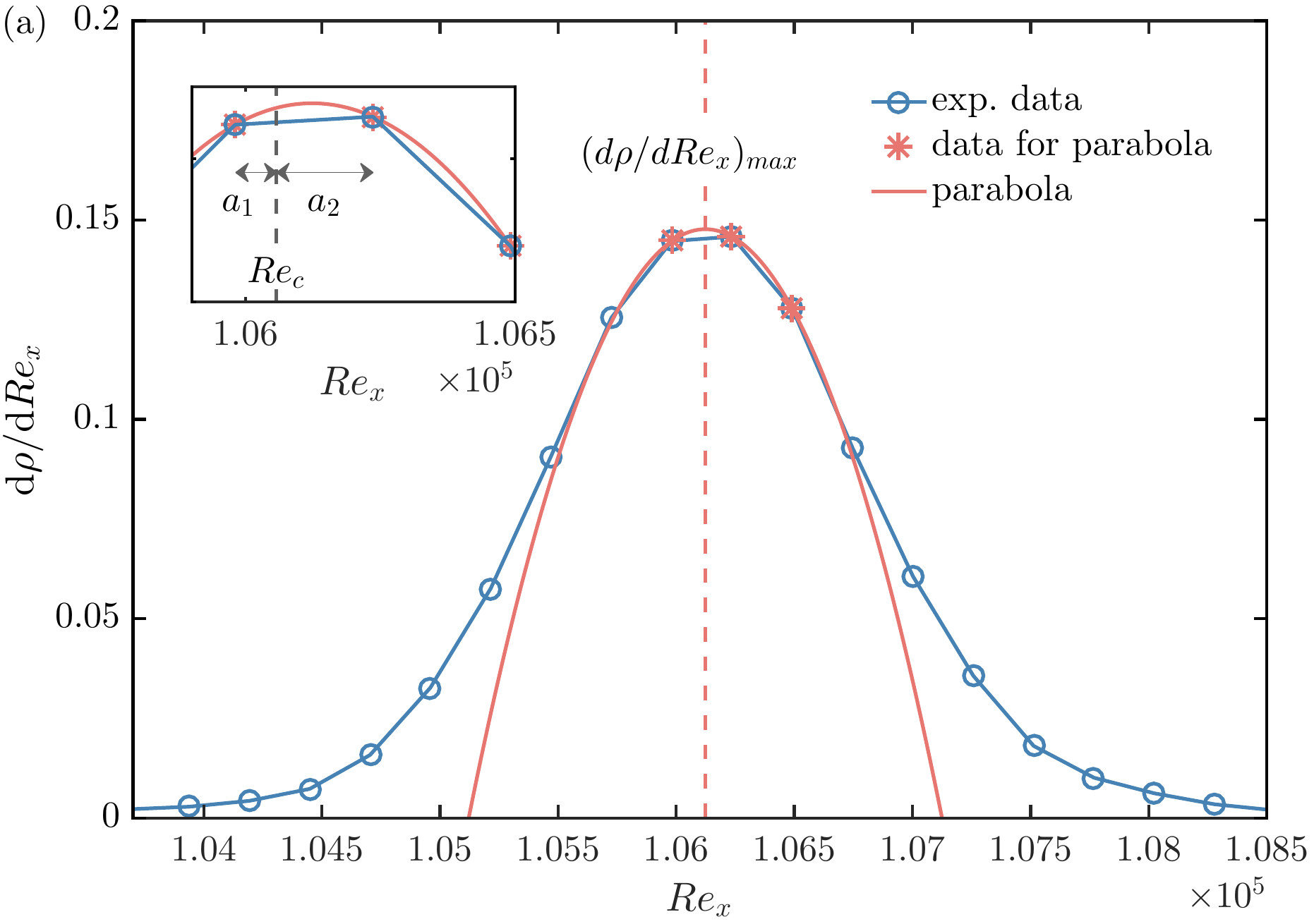}
		\includegraphics[width=0.95\columnwidth]{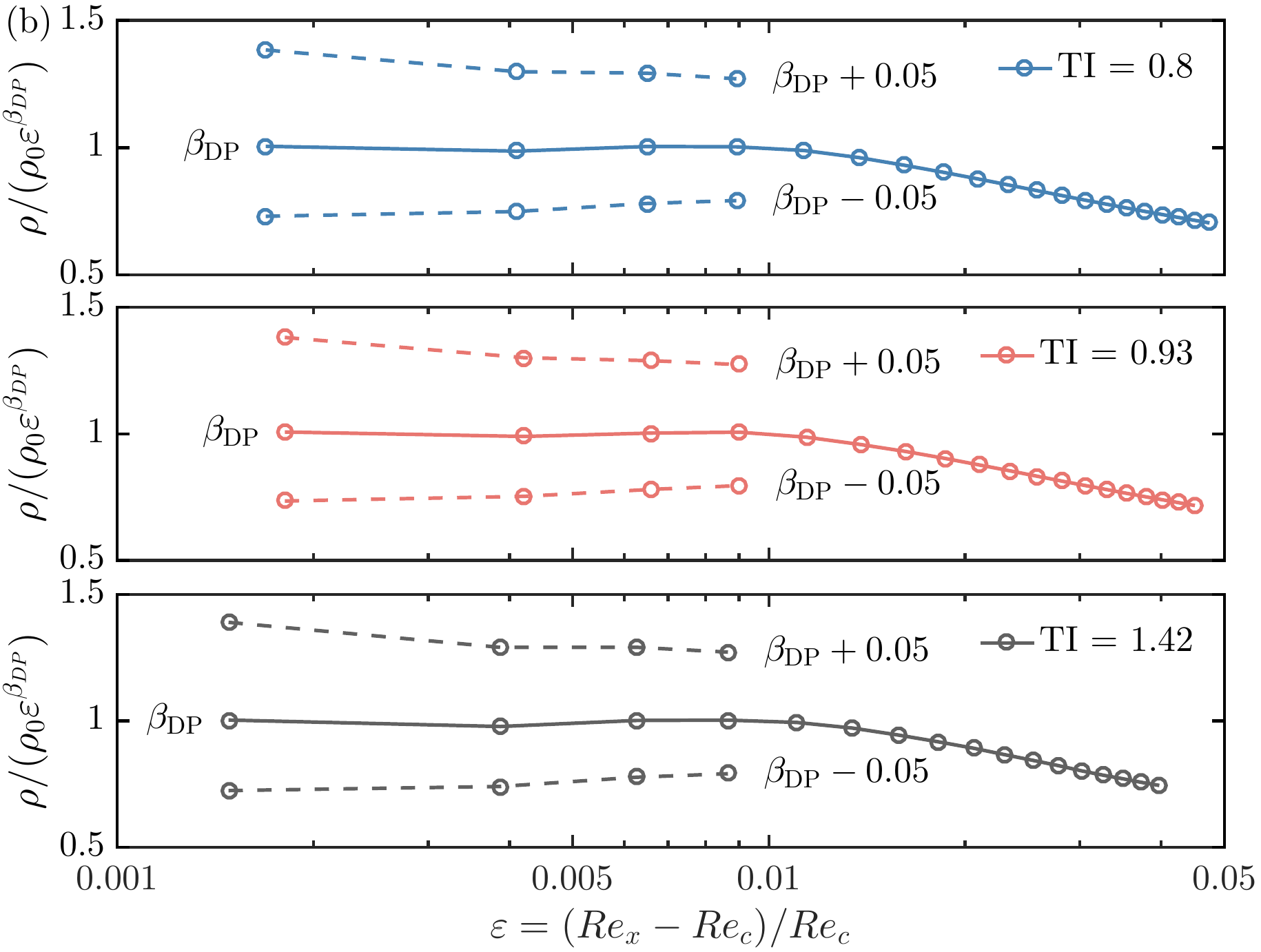}
		\caption{\protect
			\xxb{%
				(a) Illustration of how the critical Reynolds number ${\rm Re}_{\rm c}$ is determined:
				(i) One first adjusts a parabola (red line) around the three highest values (stars) of the derivative of the turbulent fraction $\frac{d\rho}{d{\rm Re_x}}$ (circles) and 
				(ii) performs a best fit expecting the critical Reynolds number ${\rm Re}_{\rm c}$ lying close to the mathematical maximum (red dashed line) of the parabola respecting the spatial resolution of the experiment. Based on this determined interval, ${\rm Re}_{\rm c}$ is estimated after Eq. (\ref{eq:beta}). The obtained ${\rm Re}_{\rm c}$ is marked in inset (gray dashed line) along with weights $a_1$ and $a_2$ used for computation of distributions of laminar cluster sizes at ${\rm Re}_{\rm c}$ (see Fig. \ref{fig06}). Here, ${\rm TI}_{\rm th}=0.8$. 
				(b) Compensated plots (after Eq. (\ref{rho_comp})) for each value of the turbulence intensity threshold: fraction of turbulent cells rescaled to the one for $1+1$D directed percolation predictions as a function of the reduced Reynolds number $\varepsilon$.
				}
				}
		\label{fig05}
	\end{center}
\end{figure}
\begin{figure*}[t]
	\begin{center}
		\includegraphics[width=0.95\columnwidth]{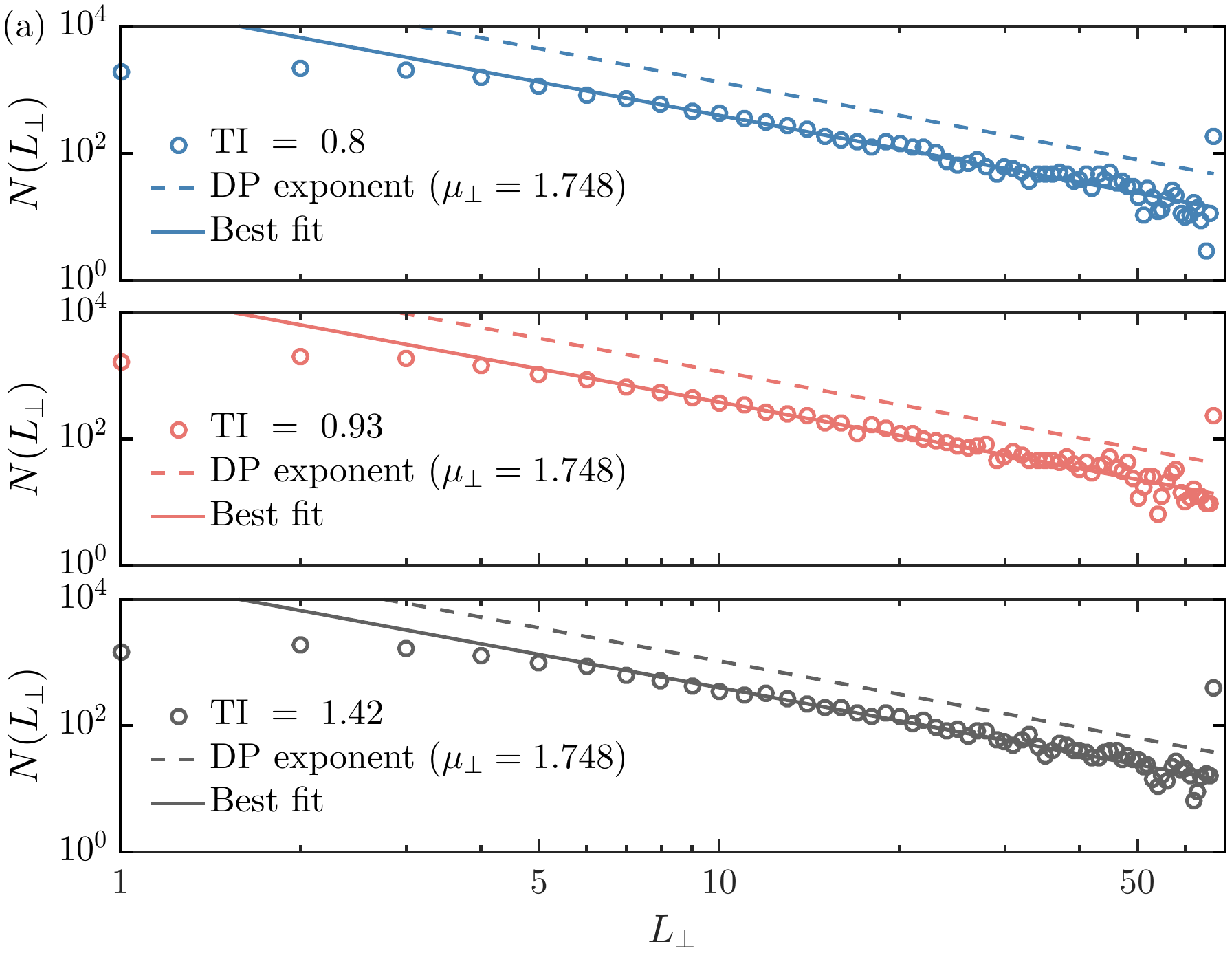}%
		\hspace{0.5cm}%
		\includegraphics[width=0.95\columnwidth]{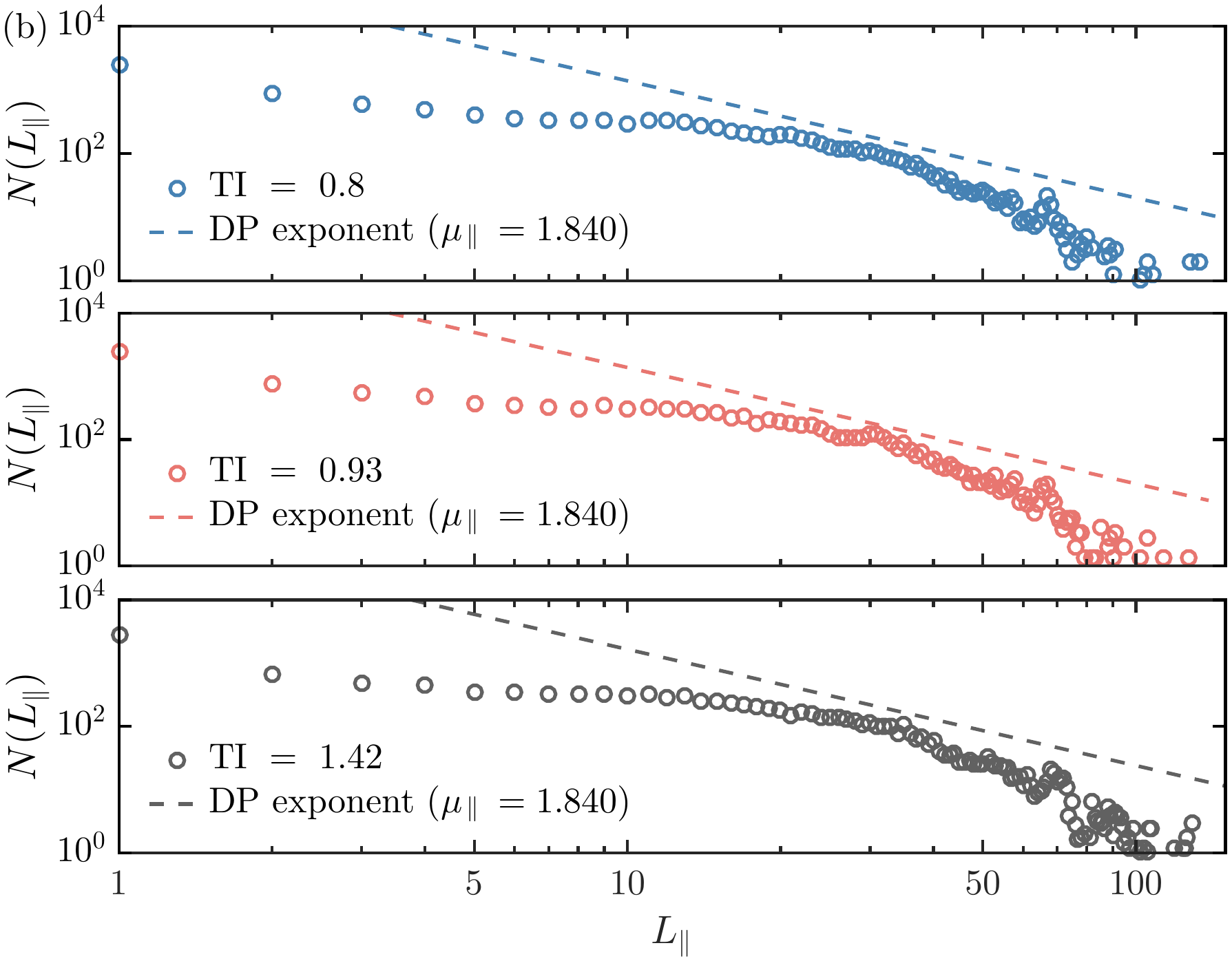}
		\caption{\protect
			\xxb{%
				(a) Size distribution of laminar clusters in space at the estimated critical Reynolds number, i.e.~consecutive laminar states when sweeping in the $y$ direction at a given time $t$, for three illustrative values of turbulence intensity thresholds and 
				(b) their corresponding size distribution of laminar clusters in time, i.e.~consecutive laminar states when sweeping in the time $t$ while keeping $y$ constant. Solid lines show the theoretical distribution predicted by the directed percolation model at the transition. For comparative purposes, the turbulence intensity thresholds values are the same as the ones in Fig.~\ref{fig04}b, where the critical value for each $\rm TI$ is given. Cluster lengths are shown in multiples of the spatial and temporal resolution respectively.}
		}
		\label{fig06}
	\end{center}
\end{figure*}

\xxb{The turbulent fraction over local Reynolds number is shown for three descriptive thresholds ${\rm TI}_{\rm th} \in [0.8,~0.93,~1.42]$ in Fig. \ref{fig04}b. Phase transition occurs within the narrow range of values of ${\rm Re}_{x}$ where an abrupt increase of the turbulent fraction is observed separating a laminar phase ($\rho \sim 0$) from a turbulent phase ($\rho\sim 1$). For the whole DP analysis, it is important to determine precisely the critical value of ${\rm Re}_{{\rm c}}$.}

\xxb{${\rm Re}_{\rm c}$ and $\beta$ can be obtained simultaneously by a best fit according to Eq. (\ref{eq:beta}). The measurement points used for the best fit are selected in two steps. At first, the derivative of the turbulent fraction with respect to the Reynolds number $\frac{d\rho}{d{\rm Re}_{x}}$ is computed, as shown in Fig. \ref{fig05}a. Considering the largest value of the derivative together with the derivative values at the two nearest measured Reynolds numbers (red stars in Fig.~\ref{fig05}a), a parabola with negative concavity is defined, whose mathematical maximum is taken as an initial estimate of the critical value. While starting the best fit at this estimated location, in step two, only measurement points are taken into account meeting $0.001 < \varepsilon < 0.01$, in accordance with the experimental spatial resolution. As shown in Fig. \ref{fig04}b, this procedure yields values of ${\rm Re}_{\rm c} = 1.061(9)\times10^5$, $\beta=0.28(4)$ for ${\rm TI}_{\rm th} = 0.8$, ${\rm Re}_{\rm c} = 1.063(6)\times10^5$, $\beta=0.28(0.05)$ for ${\rm TI}_{\rm th} = 0.93$ and ${\rm Re}_{\rm c}= 1.07(1)\times 10^5$, $\beta=0.28(5)$ for ${\rm TI}_{\rm th} = 1.42$. In comparison, the predicted theoretical value of $1+1$D directed percolation is $\beta_{DP}=0.276$.}

\xxb{In order to evaluate the validity of the determined characteristic values ${\rm Re}_{\rm c}$ and $\beta$, the rescaled turbulent fraction, $\tilde{\rho}$,  defined as
\begin{equation} \label{rho_comp}
\tilde{\rho}({\rm Re}_{x}) = 
\frac{\rho({\rm Re}_{x})}{\rho_0 \varepsilon^{\beta_{DP}}} ,
\end{equation}
is shown in Fig. \ref{fig05}b as a function of the reduced Reynolds number using the theoretical value of $\beta_{DP}$ . In this representation, scaling is readily apparent as a horizontal line equal to unity. Dashed lines also show the rescaled fraction obtained for slight deviations from the exponent, $\beta_{DP}\pm 0.05$.}

\xxb{To estimate the other two critical exponents describing the fractal dimensions, the distributions of the sizes of the laminar clusters at the critical Reynolds number have to be known. Based on our finite experimental resolution, we have no direct access to these critical distributions. To overcome this problem we take the corresponding distributions at the two Reynolds numbers close to ${\rm Re}_{\rm c}$  and sum them up in a weighted manner using the normalized inverse of their distances  $a_1$ and $a_2$ to ${\rm Re}_{\rm c}$ (cf. inset of Fig.~\ref{fig05}a).}

\xxb{While the procedure of estimating the critical size distributions
is not standard, it results in robust estimates of spatial critical exponents $\mu_{\perp}$. The distributions of cluster sizes in space and time at the critical Reynolds number are shown in Figs.~\ref{fig06}a and \ref{fig06}b, respectively. The evaluations are performed with same three threshold values ${\rm TI}_{\rm th}$ as previously.
Each best fit for the spatial fractal dimension (smallest three cluster sizes are disregarded) yields $\mu_{\perp}=1.75(9)$ and compares well with the theoretical value predicted by directed percolation, $\mu_{\perp, DP}=1.748$. For the temporal direction, experimental data is inconclusive so that a best fit is not performed. The theoretical exponent, $\mu_{\parallel, DP}=1.84$ is shown in Fig.~\ref{fig06}b as a dashed line.}

\begin{figure}[t]
  \begin{center}
    \includegraphics[clip,viewport = 30 0 600 390,width=0.95\columnwidth]{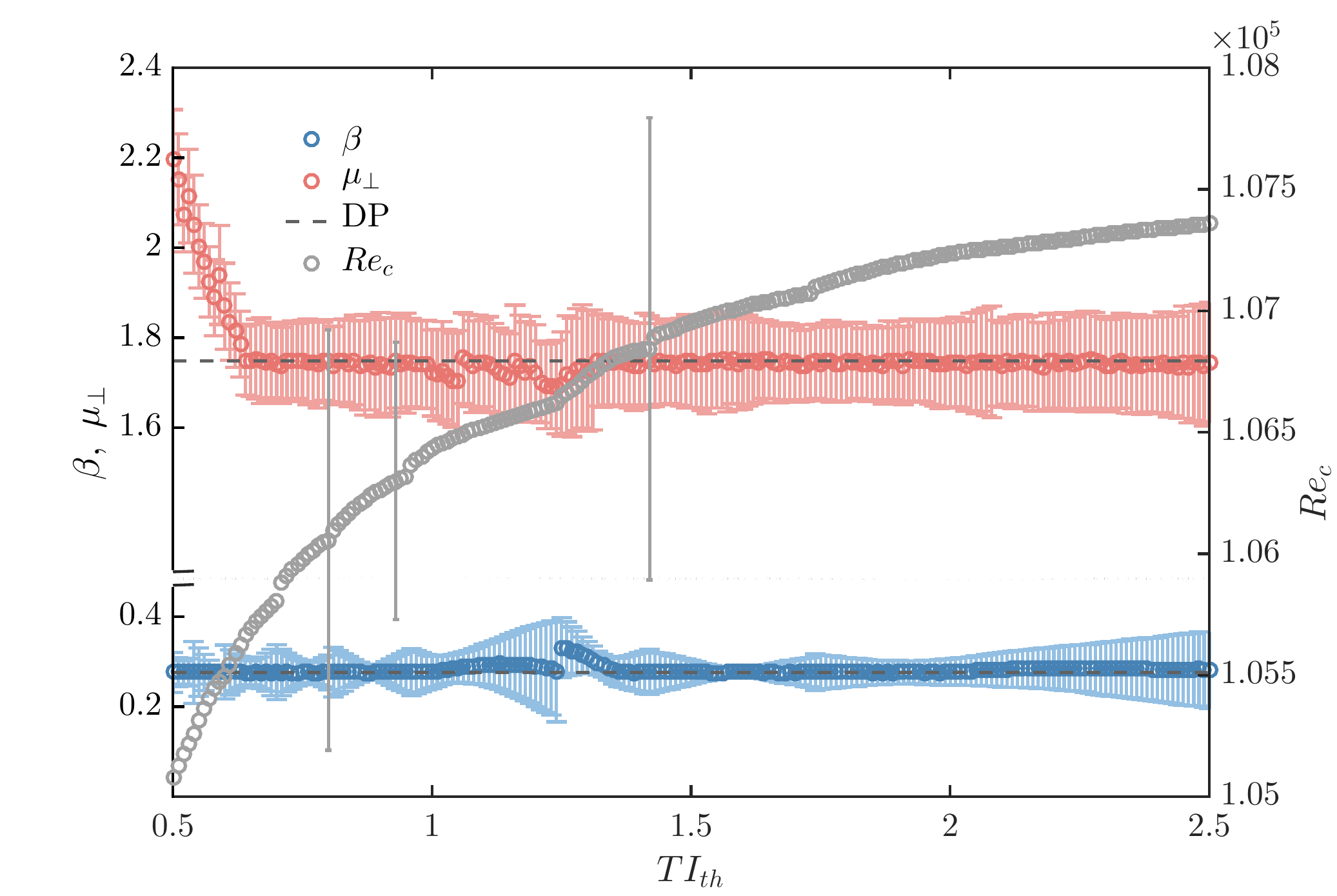}
    \caption{\protect
      \xxb{%
	Systematic analysis of the critical exponents sensitivity to the threshold of turbulence intensity imposed in the directed percolation model: the exponent $\beta$ and the transverse fractal dimension $\mu_{\perp}$. The dashed lines indicate the theoretical value of the exponents in directed percolation. For a turbulence intensity ${\rm TI}_{\rm th}>0.64$, the critical exponents of directed percolation are obtained within numerical errors (see text). Gray symbols show the dependence of the critical Reynolds number on the threshold. For the sake of clarity, only three illustrative error bars of $Re_{\rm c}$ are shown.}
	}
    \label{fig07}
  \end{center}
\end{figure}

\xxb{As a last step in our data analysis we investigate the effect of the choice of the threshold values ${\rm TI}_{\rm th}$ for this phase transition. 
Beyond the three threshold values considered up to now, Fig. \ref{fig07} shows the results for two of the three critical exponents with 95\% confidence intervals for 200 values of ${\rm TI}_{\rm th}$ covering a range of $0.5 < {\rm TI}_{\rm th} < 2.5$.  The theoretically predicted values of the critical exponents are shown by horizontal dashed lines. It is evident that the values of $\beta$ and $\mu_{\perp}$ exponents are robust against the change of the turbulence intensity threshold for ${\rm TI}_{\rm th}>0.64$.  A variation of a factor of $5$ in the turbulence intensity threshold implies a variation of less than $2\%$ of the critical Reynolds number. Thus, we conclude that the critical Reynolds number marking the onset of the LSB, depends only slightly on ${\rm TI}_{\rm th}$.}

\section{Discussion}\label{sec:discussion} 

\xxb{The main aspect of our work is to show evidence that the onset of an LSB on an airfoil is linked to directed percolation. This evidence is shown by estimated critical exponents. It is well known that this estimation is very sensitive to the choice of the critical point, or here, the critical Reynolds number. To show the quality of our estimate, we present compensated plots in Fig.~\ref{fig05}b. Another point supporting our choice of $\rm Re_c$ is that the cluster size distributions become exponentially shaped (not shown here) when cluster size distributions are computed at Reynolds numbers which differ by less than 1000 from the critical value. For the critical exponents of the turbulent fraction ($\beta$) as well as the transverse fractal dimension ($\mu_\perp$), Figs.~\ref{fig05}b, \ref{fig06}a and \ref{fig07} show a robust accordance with the predicted values of $1+1$D directed percolation. This also holds true if the confidence intervals for our error estimation are put into question. The results on longitudinal fractal dimension ($\mu_\parallel$) are inconclusive but can be taken as consistent with (or not contradicting) $1+1$D directed percolation, see Fig.~\ref{fig06}b. Particularly cluster distributions in time are known to be more difficult to estimate as these seem to suffer more from finite size effects and experimental uncertainty.}

\xxb{The obtained scaling ranges for our systems are not much more than one decade. Definitely, it would be desirable to investigate finite size scaling. In contrast to numerical investigations, we are limited by our experimental setup. Neither the spatial dynamic range of HSPIV nor the size of the wind tunnel can easily be changed. Taking also into account that estimation of power laws may lead to biased estimation and pitfalls \cite{newman}, our experimental results indicate, to the best of our belief, astonishingly consistent results with the $1+1$D directed percolation. We also see that the quality of our results are comparable with those of other groups, like the work \cite{sano2016} on directed percolation in a channel flow.}
\begin{figure}[t]
	\begin{center}
		\includegraphics[clip,viewport = 30 0 560 380,width=0.95\columnwidth]{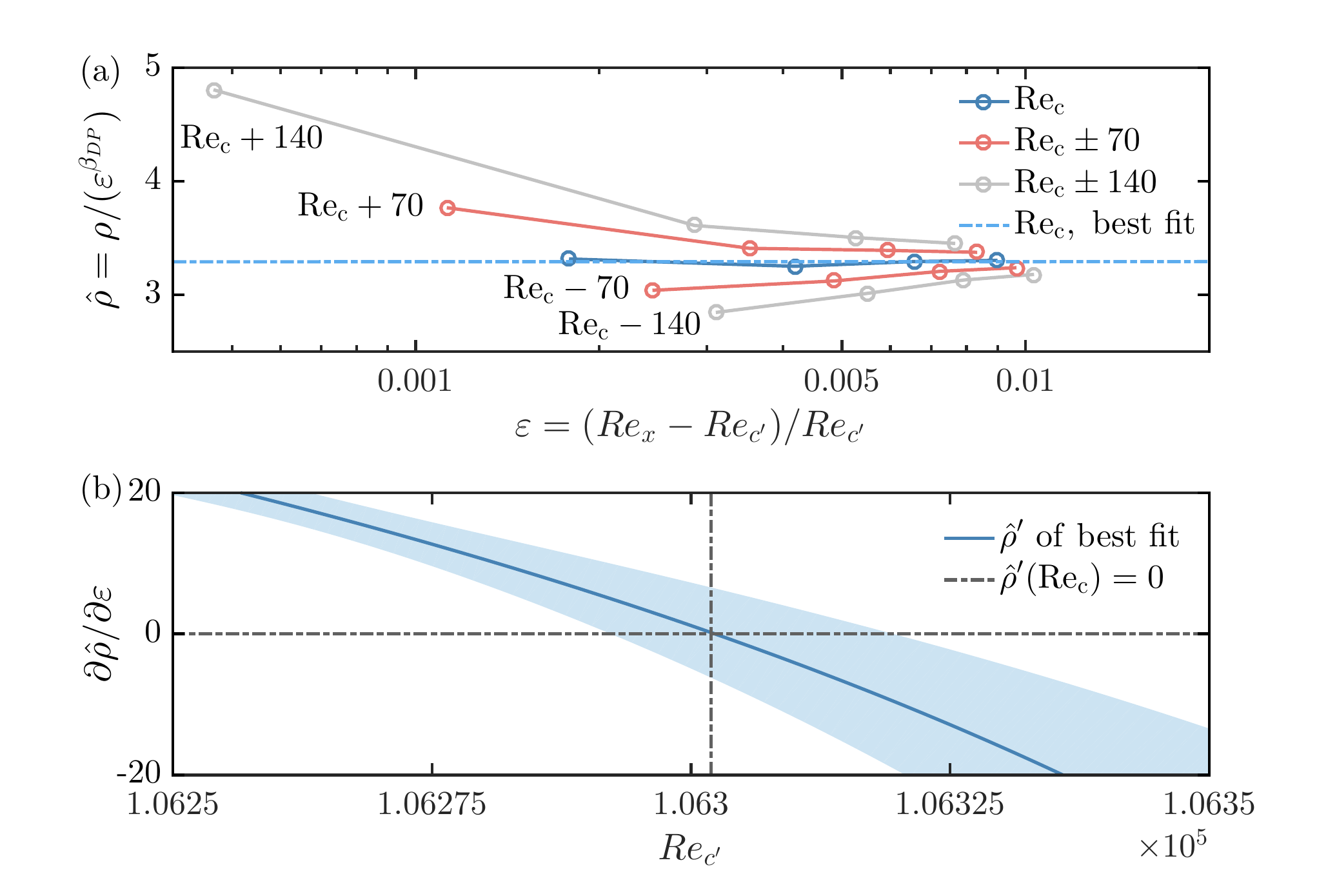}
		\caption{\protect
			\xxb{%
				(a) Rescaled turbulent fraction $\hat{\rho}$ as a function of $\varepsilon$ for a variety of possible critical Reynolds numbers, ${\rm Re}_{c^\prime} \in [{\rm Re_c},~{\rm Re_c}\pm 75,~{\rm Re_c}\pm 150]$. Where $\hat\rho$ becomes constant, ${\rm Re_c}$ can be estimated best.
				(b) Derivative of $\hat{\rho}$ as a function of ${\rm Re}_{c^\prime}$ including standard errors. The critical Reynolds number is determined by the zero-crossing of $\hat{\rho}^\prime$. The uncertainty of ${\rm Re}_{\rm c}$ is estimated by propagation of errors. 
				Here, ${\rm TI}_{\rm th}=0.93$.}
		}
		\label{fig08}
	\end{center}
\end{figure}

\xxb{Based on these findings, it is now possible to introduce an alternative procedure to determine the onset of the LSB. Under the assumption of $1+1$D DP holding true, the critical Reynolds number can be determined following the compensated representation shown in Fig.~\ref{fig05}b, but now fixing ${\beta_{DP}}$ to its theoretical value and varying  ${\rm Re}_{c^\prime}$,
\begin{equation} \label{eq:rho_comp2}
\hat{\rho}({\rm Re}_{x},{\rm Re}_{c^\prime}) = 
\frac{\rho}{\varepsilon^{\beta_{DP}}} .
\end{equation}
For a variety of possible critical Reynolds numbers, $\hat{\rho}$ is shown as a function of $\varepsilon$ and ${\rm Re}_{c^\prime}$ in Fig.~\ref{fig08}a. While increasing ${\rm Re}_{c^\prime}$, the slope, $\hat{\rho}^\prime=\partial\hat\rho/\partial{\rm \varepsilon}$, changes from positive to negative. In analogy to Fig.~\ref{fig05}b, the horizontal line in $\hat{\rho}$ represents the case where DP properties are found. Thus, from the zero-crossing of $\hat{\rho}^\prime$ the best estimation $\rm Re_{c}$ is obtained. In comparison to the estimation of the critical Reynolds number by fitting a power law about the maximum derivative of $\rho$ (see Figs.~\ref{fig04}b~and~\ref{fig05}a), the uncertainty in $\rm Re_c$  decreases by two orders of magnitude from ${\rm Re}_{\rm c} = 1.063(6)\times10^5$ to ${\rm Re}_{\rm c} = 1.06302(2)\times10^5$ for $\rm TI_{th} = 0.93$. This shows that the concept of DP enables to determine the critical Reynolds number for the LSB with very high precision. While such high precision concepts are very rare in fluid mechanical research, the introduced procedure may serve as a new benchmark.}

\section{Conclusion}\label{sec:conclusions}

\xxb{This work presents a first experimental evidence that directed percolation, as a more general concept, is also valid for practical relevant aerodynamics, namely the flow over the suction side of an airfoil. Our work has been inspired by recent achievements in fundamental turbulence research that link the onset of turbulence to directed percolation phase transition. In comparison to the flow situations investigated up to now, the flow over an airfoil changes its Reynolds number along with its stability while evolving downstream. In this sense, a new kind of spatially dependent DP is present, for which an adiabatic approximation has been anticipated.}

\xxb{Applying a bond directed percolation model to characterize transition from a laminar boundary layer into an LSB on an airfoil, one obtains values for the critical exponents consistent with those in $1+1$D directed percolation. The physical implication of this universality class indicates that the boundary layer at the LSB's onset is slender compared to the dimension of the LSB and, thus, flow instabilities cannot spread perpendicular to the surface. As an important aspect for practical applications, with the assumption of an $1+1$D directed percolation, a new method is introduced to determine the transition point into the LSB with very high precision of better than 10 in Reynolds numbers. This is for fixed TI less than 1 per mill or for all TI less than 1 per cent of the cord length. For our profile, the precision exceeds the optical resolution of $0.4~{\rm mm}$.}

\xxb{Since instabilities like the LSB have essential impact on the performance of airfoils, it is of great importance to know how and where they emerge. From the findings of this work, new directions for future applications and investigation are now open. First, in CFD, several frameworks need a model that delivers the location of transition \cite{Spalart1992} and, thus, the LSB's onset. Since the directed percolation framework is now shown to be able to retrieve a consistent determination where the LSB onset is located, it can be used to validate current transition models. Additionally, the temporal evolution of the LSB's onset parameterized by DP might be of use for unsteady low-order models \cite{Menter2015} which often couple details of the boundary layer flow with CFD approaches used for the ambient flow \cite{Spalart1997,Strelets2001}.}

\xxb{ Second, in a lot of applications, transition to turbulence is a phenomenon that needs to be controlled properly, also when taking place within an LSB. For instance, vortex generators are applied to rotor blades of wind turbines close to the location of an LSB in order to avoid aerodynamic instabilities that cause high fatigue loads. Nowadays, this is done based on efficient engineering models used in the design process of a wind turbine. As revealed by different studies \cite{Bak1999,Ward1963}, these models cannot account for nonlinear flow behavior that inherently governs the emergence of LSBs. Particularly when an LSB emerges very close to the natural onset of laminar-turbulent transition, models fail to correctly predict both phenomena. The more reliably such situations can be identified and characterized, the better vortex generators can be applied to rotor blades, resulting in more efficient wind turbine operation along with reduction of destructive aerodynamic loads.}

Third, the wind tunnel experiment here investigated considered an airfoil subjected to a constant ambient inflow. In reality, airfoils of a plane or a wind turbine are subjected to non-stationary inflows. Assuming that the non-stationarity of such flow occurs at a smaller time scale than the time scale needed for the LSB onset to take place as a percolation transition, the concept of directed percolation has the potential to derive a model for the (non-stationary) dynamics of the LSB's onset in time.

All in all, results of the present study indicate a more comprehensive significance of percolation models in fluid mechanics beyond fundamental laminar-turbulent transition phenomena.

\begin{acknowledgments}
The authors thank comments by Philipp Maa\ss, Iv\'an H\'erraez, Matthias Schramm and Bruno Eckhardt as well as technical advice from Jarek Puczylowski.
\end{acknowledgments}

%

\end{document}